\documentstyle[12pt]{article}
\def\ee{{\rm e}}

\begin{document}
\begin{titlepage}
\rightline{hep-th/9504004}
\rightline{EPHOU-95-001}
\rightline{April 1995}
\begin{center}
{\Large {\bf Multi-Instanton Effect in Two Dimensional QCD}}\\
\vspace{2.0cm}
{\large  Tetsuyuki Ochiai }\\
\vspace{0.3cm}
ochiai@particle.phys.hokudai.ac.jp  \\
\vspace{0.3cm}
{\it Department of Physics \\ Hokkaido University \\ Sapporo 060 ,
Japan }
\end{center}
\vspace{4.0cm}

\abstract{
We analyze multi-instanton sector in two dimensional U($N$) Yang-Mills theory
 on a sphere.
We obtain a contour intregral representation of the multi-instanton amplitude
and find ``neutral'' configurations of the even number instantons
 are dominant in the large $N$ limit.
Using this representation, we calculate the 1,2,3,4 bodies interactions
and the free energies for $N =3,4,5$
numerically and find that in fact the multi-instanton interaction effect
essentially contribute to the large $N$ phase transition  discovered by
Douglas and Kazakov.}
\end{titlepage}

\section{Introduction}

In the last two years, there is remarkable progress toward
understanding relation  between large $N$ QCD and string
theory.  Important contribution to this subject was made by
Gross and Taylor\cite{gt}. They showed that ``two dimensional QCD is a
string theory''. Strictly speaking, the partition function of U($N$)
Yang-Mills theory on Riemann surface $\Sigma_G$ is equivalent to that
of no fold string theory with target space  $\Sigma_G$, in other
words, sum over branched  covering maps of Riemann surface
$\Sigma_G$.  But Douglas and Kazakov discovered that on the topology of
sphere  the system has a 3rd order phase transition at $(\lambda A)_c =\pi^2$
  in the large $N$ limit
and showed  the  equivalence is  restricted within the strong coupling
region $\lambda A>\pi^2$ \cite{dk}.

Originally the phase transition was
discovered from the analysis of so called Cauchy problem in this system.
It is well known that when we calculate matrix model partition function
in the large $N$ limit, we encounter the  Cauchy problem in eigenvalue
distribution function.
In our case the partition function has the similar structure as the one of
Gaussian matrix model and we encounter a two-cut Cauchy problem
in   $\lambda A>\pi^2$. By exploring  the free energy, there is  3rd
order gap between the weak and strong coupling region.

One physical explanation of the phase transition is
the following. From the string point of view (it is view from one side i.e.
strong coupling region), covering map from sphere to sphere (which
corresponde to $1/N$ leading term of Yang-Mills free energy on sphere)
allows arbitrary but even number of branch points (2n-2) according to arbitrary
number of winding (n). Each branch point has competing  entropy (target space
area A) and energy. In additon there are
many ways to  construct world sheet from n  target spaces.
 If the entropy beat the energy , the string expansion
becomes infinite and  the phase transition occurs \cite{t}.
This scenario succeed to
explain the large $N$ phase transition qualitatively and
semi-quantitatively(it gives critical coupling  $(\lambda A)_c \simeq 11.9 $).
 But this point of view seems to be insufficient
because  it lacks global perspective and
gauge field point of view.

  Recentry, Gross and Matytsin showed that instanton
induces the phase transition \cite{gm1}.
They obtained the following results.
\begin{enumerate}
\item Yang-Mills partition function on sphere can be expressed by sum
over instantons.
\item The 0 instanton sector gives the weak coupling result.
\item In the weak coupling region the 1 instanton amplitude of charge $\pm 1$
has exponential damping factor and thus is negligigle.
\end{enumerate}
In addition, the 1 instanton amplitude of charge $\pm 1$ becomes oscilating
in the strong coupling region. But it's contribution to the free energy are
order $N^{-1/2}$ and thus negligible in the large $N$ limit.
Hence it is important to investigate the effect of the multi-instanton.

 In this paper we explore the
effect of multi-instanton and find out that there is some sort of neutrality
in the large $N$ limit. That is, ``neutral'' configurations of the
multi-instanton are dominated in this limit. We also find that
the interaction rather than the entropy  essentially
cause the phase transition. We conjecture 2 bodies force, 4 bodies force, 6
bodies force,$\ldots$ equally contribute to the large $N$ phase transition.

 The content of this paper is following.
 In  section 2  the multi-instanton amplitude and the partition
function are rewritten to  multiple contour integrals and the neutrality is
found.
 In section 3 nature of one instanton and two instantons systems are
 analized in detail.
 In section 4 numerical calculation of
 the multi-instanton amplitude and the free energy are performed.
 In appendix
using the above formulation, the multi-instanton contribution to Wilson
loop is rewritten to multiple contour integrals.

\section{ Contour integral representation and neutrality }

We begin with Migdal-Rusakov  heat kernel representation  for the
Yang-Mills  partition function on a sphere\cite{mr},
\begin{eqnarray}
    Z_{QCD}&=& \int {\cal D}A_{\mu} \exp(-\frac{N}{4\lambda}\int
                       d^{2} x \sqrt{g}{\rm tr}F_{\mu \nu}F^{\mu\nu})
                        \nonumber \\
   &=&\sum_{R} ({\rm dim} R)^{2} \exp (-\frac{\lambda A}{2N} C_{2}(R)) ,
\end{eqnarray}
where $A$ is area of the sphere,
 $R$ is irreducible representation of the gauge group U($N$) and $C_2(R)$ is
the value of the  second Casimir operator of rep $R$.
Because $R$ is characterlized by Young tableau, sum over  row lengths in
 Young tableau replaces sum over representations.
We remark that the condition that Young tableau must be stairway -like
is irrevant because  dim$R$ has form of the Van der Monde determinant and thus
has permutation symmetry.
Then the sum becomes free sum over ${\bf Z}^N $ and we can rewrite
it by the Poisson resumation formula.
We get the following expression \footnote{Hereafter we absorb
$\lambda$ into A}\footnote{A similar expression was obtained by
M. Caselle et al \cite{cdmp}. They showed that the phase
transition is due to the winding modes of ``fermion on circle''.}
\cite{w,mp,gm1}:
\begin{eqnarray}
      Z_{QCD}&=& \ee^{\frac{A}{24}(N^2-1)}({N\over A})^{N^2}
          \sum_{\{m_i\} \in {\bf Z}^N}
          \ee^{-\frac{2 \pi^2 N}{A} \sum _{i=1}^{N } m_{i}^{2}} w(\{m\}) ,\\
      w(\{m\})&\equiv &\int_{-\infty}^{\infty} \prod_{i=1}^{N} dy_{i}
                          \prod_{i<j}^{N} (y_{ij}^{2} -4\pi^{2} m_{ij}^{2} )
                          \ee^{- \frac{N}{2A} \sum_{i=1}^{N} y_{i}^{2} }
                           \nonumber \\
         &=&\int_{-\infty}^{\infty} \prod_{i=1}^{N} dy_{i}
                  \Delta (y_{i}+2\pi m_{i})\Delta (y_{i}-2\pi m_{i})
                  \ee^{- \frac{N}{2A} \sum_{i=1}^{N} y_{i}^{2} }
\end{eqnarray}
where $y_{ij}=y_i-y_j$ and $m_{ij}=m_i-m_j$.
In \cite{w,mp,gm1} it was showed that
$\{m\}$  corresponde to all Euclidean classical solutions
(which we call instanton) up to gauge transformation and  has nonperturbative
 effect with respect
to both $1/N$ and $\lambda$. Hereafter we call $m_{i}$
 instanton charge and call number of nonzero $m_{i}$'s  instanton number.
The $ w(\{m\}) $ looks like Gaussian matrix model but there is a
deformation in the Van der Monde determinant $\Delta$.
In this section we rewrite $w(\{m\})$  using the method of
ortho-polynomial (in this case Hermite polynomial) and  obtain
a new multiple contour integrals representation.
 The new representation makes clear the role of the multi-instanton which
was not known.

\subsection{ 0 and 1 instanton sector }

    In the case of 0 instanton (i.e all $m_{i}=0$),$w(\{m\})$ is itself
parition function of  Gaussian matrix model and has the following
form:
\begin{equation}
      w(0) =\int_{-\infty}^{\infty} \prod_{i=1}^{N} dy_{i}
                          \Delta^{2}(y)
                          \ee^{- \frac{N}{2A} \sum_{i=1}^{N} y_{i}^{2} }
               \equiv Z_{G}(\frac{N}{2A}) =c_N ({N\over 2A})^{-{N^2\over 2}} .
\end{equation}

 In the case of 1 instanton (i.e. $m_{i}=m\delta_{ik}$),
Using the property of the Van der Monde determinant and the Taylor series
expansion of the  Hermite polynomial, we obtain,
\begin{eqnarray}
 w(m) &=& \int_{-\infty}^{\infty}
           \prod_{i=1}^{N} dy_{i}
           \Delta (y_{i}+2\pi m \delta_{ik}) \Delta (y_{i}-2\pi m \delta_{ik})
           \ee^{- \frac{N}{2A} \sum_{i=1}^{N} y_{i}^{2} }  \nonumber \\
      &=& (N-1)! \sum_{n=0}^{N-1} h_{0} \cdots \check{h_{n}} \cdots
                h_{N-1} \int_{-\infty}^{\infty} dy_{k} P_{n}(y_{k}+2\pi m)
                P_{n}(y_{k}-2\pi m)  \ee^{- \frac{N}{2A} y_{k}^{2}}
                    \nonumber \\
      &=& \frac{1}{N} Z_{G}(\frac{N}{2A})
                 \sum_{n=0}^{N-1}\sum_{l=0}^{n}
                ({}_{n} C_{l} )^{2} (-4\pi^2 m^2)^{n-l} \frac{h_{l}}{h_{n}}
                 \nonumber \\
      &=& \frac{1}{N} Z_{G}(\frac{N}{2A}) \oint \frac{dt}{2 \pi i}
              \ee^{-\frac{ 4 \pi^2 m^2 N t}{A}} (1+\frac{1}{t})^{N} ,
\end{eqnarray}
where
\begin{equation}
P_n (x)=\frac{1}{2^n (\frac{N}{2A})^{\frac{n}{2}}}H_n
(\sqrt{\frac{N}{2A}}x)
\end{equation}
and
\begin{equation}
\int_{-\infty}^{\infty} dy P_{n}(y)
                P_{m}(y)  \ee^{- \frac{N}{2A} y^2} =h_{n}\delta_{nm}
                =\sqrt{2\pi}(\frac{A}{N})^{n+\frac{1}{2}} n! \delta_{nm} .
\end{equation}
This contour integral representation for the 1 instanton amplitude
was obtained in \cite{gm1} from another
aspect. They
calculated it using large $N$ saddle point method and  found it to have
 behavior change from exponential damping to oscilating at
 $A=m^2 \pi^2$. The detail is given in section 3.

\subsection{ multi-instanton sector }

  In the same way, we can rewrite  multi-instanton amplitudes to
multiple contour integrals:
\begin{eqnarray}
& & \qquad w(\{m\}) \nonumber \\
&=& \int_{-\infty}^{\infty}
              \prod_{i=1}^{N} dy_{i}
              \Delta (y_{i} +2\pi m_{i} )
              \Delta (y_{i}-2\pi m_{i} )
              \ee^{- \frac{N}{2A} \sum_{i=1}^{N} y_{i}^{2} } \nonumber \\
&=& \sum_{\mu \in S_{N}}  {\rm sgn} \mu \sum_{\sigma \in
              S_{N}}
              \int_{-\infty}^{\infty} \prod_{i=1}^{N} dy_{i}
              \prod_{i=1}^{N} P_{\sigma (i)} (y_{i} +2\pi m_{i} )
              P_{\mu \circ \sigma (i)} (y_{i} -2\pi m_{i} )
               \ee^{ - \frac{N}{2A} \sum_{i=1}^{N} y_{i}^{2}  } ,
\end{eqnarray}
where $S_N$ is the permutation group on $N$ object.

 According to number of  nonzero  m's , i.e. number of instantons, the
following analysis is  different. Assume number of  nonzero  m's
is k, using permutation symmetry, nonzero  m's are driven to
$m_{i}$'s from i=1 to i=k. Then we have,
\begin{eqnarray}
 w(\{m\}) &=& \sum_{\mu \in S_{N}} {\rm sgn} \mu \sum_{\sigma \in S_{N}}
              \prod_{i=k+1}^{N} h_{\sigma ( i)}
               \delta_{\sigma ( i), \mu \circ \sigma (i) }  \nonumber \\
  & & \qquad \times \prod_{i=1}^{k} \int_{-\infty}^{\infty} dy_{i}
                P_{\sigma (i)} (y_{i} +2\pi m_{i} )
                P_{\mu \circ \sigma (i)} (y_{i} -2\pi m_{i} )
               \ee^{- \frac{N}{2A}  y_{i}^{2} } \nonumber \\
  &=& (N-k)! \sum_{\mu \in S_{k}} {\rm sgn} \mu
                  \sum_{a_{1} \neq \cdots \neq a_{k}}
                  h_{0} \cdots \check{ h_{a_{1}}}  \cdots
                  \check{h_{a_{k}}}\cdots h_{N-1}    \nonumber  \\
  & & \qquad \times
\prod_{i=1}^{k} \int_{-\infty}^{\infty} dy_{i}
                  P_{a_{i}} (y_{i}+2\pi m_{i} )
                  P_{\mu(a_{i})} (y_{i} -2\pi m_{i} )
                  \ee^{- \frac{N}{2A}  y_{i}^{2}  } .
\end{eqnarray}
In the same way as 1 instanton sector, using the  Taylor series expansion of
the Hermite polynomial, we obtain for the above multiple integrals,
\begin{eqnarray}
& & \prod_{i=1}^{k} \int_{-\infty}^{\infty} dy_{i}
                  P_{a_{i}} (y_{i}+2\pi m_{i} )
                  P_{\mu(a_{i})} (y_{i} -2\pi m_{i} )
                  \ee^{- \frac{N}{2A}  y_{i}^{2} }
                   \times (h_{a_{1}} \cdots h_{a_{k}})^{-1} \nonumber \\
&=& \prod_{l=1}^{k} \sum_{i_{l}=0}^{{\rm min} (a_{l},\mu (a_{l}))}
                 \frac{\,_{\mu (a_{l})} C_{i_{l}}}{(a_{l}-i_{l})!}
                (\frac{2\pi m N}{A})^{a_{l}-i_{l}}
                (-2 \pi m)^{\mu (a_{l})-i_{l}} .
\end{eqnarray}
The each series in the above equation can be expressed as
a contour integral by the following
transformation formula.
\begin{equation}
\sum_{i=0}^{{\rm min}(a,b)}
                 \frac{\,_{b} C_{i}}{(a-i)!}
                (\alpha)^{a-i}  (\beta)^{b-i}
      = \oint \frac{ dt }{2\pi i} \ee^{\alpha \beta t}
                 \frac{1}{t} (\frac{1}{\beta t})^{a}
                 (\beta (t+1) )^{b} ,
\end{equation}
where the contour of $t$ encircles the origin  counterclockwise. Using
this formula, we obtain $w(\{m\})$ for k instantons,
\begin{eqnarray}
w(\{ m \} )&=& \frac{Z_{G}(\frac{N}{2A})}{N(N-1)\cdots (N-k+1)}
                  \sum_{a_{1} \neq \cdots \neq a_{k}}^{N-1}
                   \sum_{\mu \in S_{k}} {\rm sgn} \mu
                   \oint \frac{ dt_{1} }{2\pi i} \cdots
                   \oint \frac{ dt_{k} }{2\pi i}
                   \frac{1}{t_{1} \cdots t_{k}}  \nonumber \\
& & \qquad\times\ee^{-\frac{ 4 \pi^2 N}{A} (m_{1}^2  t_{1}+\cdots+
                    m_{k}^2 t_{k} ) }
                    (\frac{m_{\mu (1)}}{m_{1}}
                    \frac{1+t_{\mu (1)}}{t_{1}})^{a_{1}}
                    \cdots (\frac{m_{\mu (k)}}{m_{k}}
                    \frac{1+t_{\mu (k)}}{t_{k}})^{a_{k}} .
\end{eqnarray}
We remark the configurations such as $a_1=a_2 $ do not affect the above
equation. Hence we can drive $ \sum_{a_{1} \neq \cdots \neq a_{k}}^{N-1}$
to independent sum $\sum_{a_1}^{N-1}\cdots \sum_{a_k}^{N-1}$.
 In that expression there
are many pole free terms, then ,
after eliminating pole free terms, this integrals finally become
\begin{eqnarray}
w(\{m\})&=& \frac{Z_{G} (\frac{N}{2A})}
                     {N(N-1)\cdots (N-k+1)}
                   \oint \frac{ dt_{1} }{2\pi i} \cdots
                   \oint \frac{ dt_{k} }{2\pi i}  \nonumber \\
        & & \times \ee^{-\frac{ 4 \pi^2 N }{A}(m_{1}^2  t_{1}+\cdots+
                    m_{k}^2 t_{k} ) }
                  (1+\frac{1}{t_{1}})^{N} \cdots (1+\frac{1}{t_{k}})^{N}
                   \nonumber \\
        & &\times \sum_{\mu \in S_{k}} {\rm sgn} \mu
                   \frac{1}{\frac{m_{\mu (1)}}{m_{1}} (1+t_{\mu (1)})-t_{1}}
                   \cdots
                   \frac{1}{\frac{m_{\mu (k)}}{m_{k}} (1+t_{\mu (k)})-t_{k}} .
\end{eqnarray}
The last sum is just determinant of matrix $M$ which has $(ij)$ element
\begin{equation}
 M_{ij} \equiv  \frac{1}{\frac{m_{j}}{m_{i}} (1+t_{j})-t_{i}} .
\end{equation}
It's diagonal element is 1 and this determinant has all information about the
interaction between instantons.

\subsection{partition function}
We can rewrite the partition function as another contour integral.
For this purpose, we must expand  $w(\{m\})$ by number of instantons
 and classify elements of the permutation group $S_{k}$ into
the conjugacy classes. By symmetry argument, the contribution from different
elements belonging to same conjugacy class is same. Then we obtain,
\begin{eqnarray}
Z_{QCD}&=& Z^{{\rm weak}}
     \sum_{n=0}^{N} \sum_{m_{1},\ldots ,m_{n} \neq 0}
      {\rm e}^{-\frac{ 2 \pi^2 N }{A}(m_{1}^2 +\cdots+ m_{n}^2) }
      \oint \frac{ dt_{1} }{2\pi i} \cdots \oint \frac{ dt_{n} }{2\pi i}
      {\rm e}^{-N(\Phi_{m_{1}}(t_{1})+\cdots+ \Phi_{m_{n}}(t_{n})) }
        \nonumber \\
      & & \qquad\qquad \times \sum_{{\rm conj. class}} {\rm sgn}[\sigma ]
         T[\sigma ]
       M_{1\sigma_{1}} M_{2\sigma_{2}} \cdots M_{n\sigma_{n}} ,
\end{eqnarray}
where
\begin{eqnarray}
Z^{{\rm weak}}&\equiv & \ee^{\frac{A}{24}(N^2-1)}({N\over
       A})^{N^2}Z_{G} (\frac{N}{2A})={\rm const} \, \ee^{N^2 (\frac{A}{24}-
       \frac{1}{2}\log A ) - \frac{A}{24} } ,\\
\Phi_{m}(t) &\equiv & \frac{ 4 \pi^2 m^2 t}{A}-\log(1+\frac{1}{t}) .
\end{eqnarray}
$Z^{{\rm weak}}$ is the partiton function in the weak coupling region
and it has simple $A$
dependence.
We assume $\sigma $ has cycle structure $[1^{\sigma_{1}}\cdots
n^{\sigma_{n}}]  (\sigma_{1} +\cdots +n\sigma_{n} =n)$ and define $T[\sigma ]$
as number of elements in the conjugacy class which  $\sigma $ belongs to,
\begin{eqnarray}
     T[\sigma ] &=& \frac{n!}{1^{\sigma_{1}} \sigma_{1} ! \cdots
                               n^{\sigma_{n}} \sigma_{n} ! } , \\
     {\rm sgn} \sigma &=& (-)^{\sigma_{2}+\sigma_{4}+ \cdots} , \\
     \sum_{{\rm conj  class}} &=& \sum_{\sigma_{1},\ldots ,\sigma_{n}=0
                                                   \atop
                                         \sigma_{1} +\cdots +n\sigma_{n}
=n}  .
\end{eqnarray}
We use (12345)(67) type element as the representative element of the
conjugacy class.
 Because the multiple integrals factorize according to the cycle structure
of $\sigma$, contribution from the multi-instanton are exponentiated with
a constraint $( \sigma_{1} +\cdots +n\sigma_{n} =n) $ which is expressed by
another contour integral. We get the following result:
\begin{equation}
Z_{QCD}=Z^{{\rm weak}} \oint \frac{ dz }{2\pi i}
     \frac{1}{z^{N+1}}
     \frac{1}{1-z} \exp(\sum_{j=1}^{N} \frac{(-)^{j-1}}{j} z^{j}
     \alpha_{j} ) ,
\end{equation}
where
\begin{eqnarray}
\alpha_{j}&\equiv & \sum_{m_{1},\ldots,m_{j}\neq 0}
      {\rm e} ^{-\frac{ 2 \pi^2 N }{A}(m_{1}^2 +\cdots+ m_{j}^2) }
      \oint \frac{ dt_{1} }{2\pi i} \cdots \oint \frac{ dt_{j} }{2\pi i}
      {\rm e}^{-N(\Phi_{m_{1}}(t_{1})+\cdots+ \Phi_{m_{j}}(t_{j})) }
      \nonumber \\
      & & \qquad\qquad\qquad \times M_{12} M_{23} \cdots M_{j1}
\end{eqnarray}
is the ``connected'' amplitude of the $j$ instantons. \\

Some remark should be noted.
In the large $N$ limit, the multiple contour integrals for
$w(\{m\})$ or $\alpha_{j}$ are dominated by the saddle points.
Fortunately, it is no need for  solving complicated saddle point equations
 in our case. The saddle point equations are
decoupled and the solutions of the equations are copies of  the saddle point
found by Gross and Matytsin. Then if we neglect the interaction term
$M_{12} M_{23} \cdots M_{j1}$, $\alpha_j$ becomes factorized such as
$(\alpha_1)^j$ and the  behavior change from
exponential damping to oscilating  at $A=\pi^2 m^2 $ observed in
\cite{gm1} still hold for the multi-instanton amplitude.
But there is a remarkable feature in  the interaction term of
the multi-instanton
sector.

 We point out that in the region of $A>\pi^2$  special charge
configurations of the multi-instanton exist.  If we consider the following
configurations in $\alpha_{j} $ (even j):
\begin{equation}
m_{1}=-m_{2}=m_{3}=-m_{4}=\cdots =m ,
\end{equation}
the large $N$ saddle points for any $t_{i}$ are same and satisfy the next
relation:
\begin{equation}
t_{\pm}=\frac{-1\pm \sqrt{1-\frac{A}{\pi^2 m^2}}}{2} \quad
, \quad t_{+}+t_{-}+1=0 .
\end{equation}
When $A>m^2 \pi^2$, both saddle points must be selected. Moreover when
 $t_{i}=t_{\pm}$ and $t_{i+1}=t_{\mp}$,
all $M_{i i+1}$'s have pole and usual $N^{-1/2}$ power law (Gauss integral
factor) must be changed.
  Toy example of such phenomena is the following:
\begin{equation}
\oint \frac{dt}{2\pi i} \exp(-N\phi (t)) \frac{1}{(t-t_{c})^{n}} ,
\end{equation}
where $t_{c}$ is saddle point of $\phi (t)$.
We can approximate this to a principal value integaral along the steepest
descent line and a contour integral along small semi circle around the  saddle
point and obtain finite value with order $N^{\frac{n-1}{2}}
\exp(-N\phi (t_c))$.
We call this type of problem ``singular saddle point''.
Same power change occurs in  the multi-instanton sector.
The configuration that all
$M_{i i+1}$ have pole is
dominant contribution with respect to $N$ in the strong coupling region.
We call this phenomena ``large $N$ neutrality''.
By ``neutrality'', we would not intend general neutral configuration which
satisfy $\sum_{i=1}^N m_i =0$, but intend that all instantons have equal
 absolute charge and number of positive charge instantons and number of
negative charge instantons are same. Hence by definition ``neutrality''
 in $\alpha_j $ with odd j is ruled out. Bellow we will see $\alpha_j $
 with odd j is in fact negligible in the large $N$ limit.

\section{One body and two bodies effective action}

Below we will see generel 2n bodies interactions equally contribute to
the phase
transition. Even so, it would be worthwhile to analyze familiar two bodies
interaction  between instantons in detail.
We find that the interaction between instantons is sensitive to A and m.

\subsection{one body effective action }
First we consider the contribution of one instanton with charge $m$ to the
partition function.
In the large $N$ limit, the saddle point equation $\Phi_m^{\prime}(t)=0$
 has two solution
$t_{\pm}=\frac{-1\pm \sqrt{1-A/\pi^2 m^2}}{2} $.
We see quite different behavior according to $A {> \atop <} \pi^2 m^2$.

\subsubsection{$m^2 > A/\pi^2$}
Both saddle points  exist on negative real axis and satisfy
\begin{equation}
\Phi_{m}(t_{-})< \Phi_{m}(t_{+}) .
\end{equation}
Neverthless only $t_{+}$ saddle point is selected
since $t_{-}$ steepest decent line is not consistent with the original
contour. Hence we obtain in the large $N$ limit ,
\begin{equation}
w(m)\simeq (-)
\frac{1}{\sqrt{2\pi N |\Phi_{m}^{(2)}(t_{+})|}}
                     {\rm e}^{-N\Phi_{m}(t_{+})} .
\end{equation}
If we define $\gamma (x)$ as \cite{gm1}
\begin{equation}
\gamma (x)\equiv
\sqrt{1-x}-\frac{x}{2}\log\frac{1+\sqrt{1-x}}{1-\sqrt{1-x}} ,
\end{equation}
we obtain
\begin{equation}
Z_{1 {\rm inst}} \equiv \ee^{-\frac{2\pi^2 m^2 N}{A}} w(m)=
\frac{(-)^{N-1}}{\sqrt{2\pi N \frac{16\pi^4m^4}{A^2}
\sqrt{1-\frac{A}{\pi^2m^2}}}}
{\rm e}^{-\frac{2\pi^2 m^2N}{A}
                   \gamma(\frac{A}{\pi^2m^2})} .
\end{equation}
Since  $\gamma (x)$ is positive real for $x<1$ but pure imaginary for
$x>1$ , self energy term survives in this  region but disappears at
$A= m^2 \pi^2$.

\subsubsection{ $m^2 < A/\pi^2$  }
In this case  both saddle points are selected because $t_{\pm}$ steepst
descent lines are consisitent with original contour
and satisfy
\begin{equation}
{\rm Re}[\Phi_{m}(t_{+})]={\rm Re}[\Phi_{m}(t_{-})] .
\end{equation}
Therefore, we get
\begin{equation}
Z_{1 {\rm inst}}=
\frac{(-)^N}{\sqrt{2\pi N \frac{16\pi^4
m^4}{A^2} \sqrt{\frac{A}{\pi^2 m^2}-1}}}
  ({\rm e}^{i\frac{3}{4}\pi
                 -\frac{2\pi^2 m^2N}{A}
                   \gamma(\frac{A}{\pi^2m^2})}
  +{\rm e}^{-i\frac{3}{4}\pi
                 +\frac{2\pi^2 m^2N}{A}
                   \gamma(\frac{A}{\pi^2m^2})}) .
\end{equation}
Since $\gamma (x) $ are pure imaginary in this region, the above equation
has order $ N^{-1/2}$ in the large N limit.

\subsection{two bodies effective action}
Next we consider the contribution of two instantons with chage $m$ and $m'$
 to the
partition function in the large $N$ limit.

\subsubsection{ $ m^2 \ge m'^2 > A/\pi^2$ }

Straight forward application of large N saddle point method gives
\begin{equation}
w(m,m')=
        \frac{1}{\sqrt{2\pi N |\Phi_{m}^{(2)}(t_{+})|}}
       \frac{1}{\sqrt{2\pi N |\Phi_{m'}^{(2)}(s_{+})|}}
      {\rm e}^{-N(\Phi_{m}(t_{+})+\Phi_{m'}(s_{+}))}
      {\rm det}M(t_{+},s_{+}) ,
\end{equation}
where $t_{\pm}=\frac{-1\pm \sqrt{1-A/\pi^2 m^2}}{2}$,
$s_{\pm}=\frac{-1\pm \sqrt{1-A/\pi^2 m'^2}}{2} $ and
\begin{equation}
{\rm det}M(t,s) =1-\frac{1}{(\frac{m'}{m}(1+s)-t)
                                (\frac{m}{m'}(1+t)-s)} .
\end{equation}
As a further approximation  we assume $m'^2 \gg A/\pi^2$, then we
obtain
\begin{eqnarray}
Z_{2 {\rm inst}}&\equiv & \ee^{-\frac{2\pi^2 N}{A}(m^2+m'^2) }w(m,m')
                                      \nonumber \\
                &\sim & \frac{1}{N}\ee^{-2N(q^2+q'^2)
                                 +N(\log{q^2}+\log{q'^2})+\log(q-q')^2} ,
\end{eqnarray}
where $q\equiv \sqrt{\pi^2/A}\, m$ is effective charge and
 we used the following expansion formula of $\gamma$ around $x=0$:
\begin{equation}
\gamma (x)= 1+\frac{x}{2}\log (\frac{x}{4})-\frac{x}{2}+\sum_{n=1}^{\infty}
             \frac{(2n-1)!!}{n!(n+1)!}(\frac{x}{2})^{n+1} .
\end{equation}
We note that the interaction term
is next leading order in $1/N$ and the self energy term strongly suppresses
this configuration.

\subsubsection{ $ m^2 \leq m'^2 < A/\pi^2 $ and  $m \neq -m' $ }
In the same way we obtain,
\begin{eqnarray}
w(m,m') &=&
       \frac{1}{\sqrt{2\pi N |\Phi_{m}^{(2)}(t_{+})|}}
       \frac{1}{\sqrt{2\pi N |\Phi_{m'}^{(2)}(s_{+})|}}
      {\rm e}^{i\frac{3}{2}\pi
                   -N(\Phi_{m}(t_{+})+\Phi_{m'}(s_{+}))}
      {\rm det}M(t_{+},s_{+}) \nonumber \\
     &+&  \frac{1}{\sqrt{2\pi N |\Phi_{m}^{(2)}(t_{-})|}}
       \frac{1}{\sqrt{2\pi N |\Phi_{m'}^{(2)}(s_{+})|}}
      {\rm e}^{-N(\Phi_{m}(t_{-})+\Phi_{m'}(s_{+}))}
      {\rm det}M(t_{-},s_{+}) \nonumber \\
     &+&  \frac{1}{\sqrt{2\pi N |\Phi_{m}^{(2)}(t_{+})|}}
       \frac{1}{\sqrt{2\pi N |\Phi_{m'}^{(2)}(s_{-})|}}
      {\rm e}^{-N(\Phi_{m}(t_{+})+\Phi_{m'}(s_{-}))}
       {\rm det}M(t_{+},s_{-})  \\
     &+& \frac{1}{\sqrt{2\pi N |\Phi_{m}^{(2)}(t_{-})|}}
       \frac{1}{\sqrt{2\pi N |\Phi_{m'}^{(2)}(s_{-})|}}
       {\rm e}^{-i\frac{3}{2}\pi
                   -N(\Phi_{m}(t_{-})+\Phi_{m'}(s_{-}))}
       {\rm det}M(t_{-},s_{-}). \nonumber
\end{eqnarray}
If we further assume $m'^2 \ll A/\pi^2 $, then we obtain
\begin{equation}
Z_{2 {\rm inst}} \sim \frac{1}{N} [\theta(qq')+
                         \theta(-qq') {\rm e}^{\log(q-q')^2-\log(q+q')^2} ] ,
\end{equation}
where we used the following property of $\gamma$ :
\begin{equation}
\gamma (x)\simeq -\frac{i\pi}{2}x \qquad x\gg 1 .
\end{equation}
We note that in this region the self energy  term almost disappears and
another interaction term $\log(q+q')^2 $ is added.

\subsubsection{ $ m^2 < A/\pi^2 $  and  $ m=-m' $  }
As is alreasdy noted, in this case we encounter the  singular
saddle point.
\begin{equation}
w(m,-m )=\oint \frac{dt}{2\pi i}\oint \frac{ds}{2\pi i}
                   {\rm e}^{-N(\Phi_{m}(t)+\Phi_{m'}(s))}
                   (1-\frac{1}{(1+t+s)^2} ) .
\end{equation}
Since the 1st term of the determinant has no singularity at saddle
points, we can neglect this term and obtain
\begin{equation}
w(m,-m)\simeq -N^2 \int_{0}^{\infty} dz z {\rm e}^{-Nz} f(z)^2 ,
\end{equation}
where
\begin{equation}
f(z)\equiv  \oint \frac{dt}{2\pi i}
          {\rm e}^{-(\frac{4 \pi^2 m^2} {A}+z) N t}(1+\frac{1}{t})^{N} .
\end{equation}
{}From above equation $f(z)$ is essntially $\alpha_1$ with minor changes.
If the $z$ is above $4- 4 \pi^2 m^2 /A$ ,
$\exp(-Nz/2)f(z)$ has the behavior of exponential damp. Hence we can
restict the z integration range up to $4-4 \pi^2 m^2/A$.
In addition, $\exp (-Nz) f(z)^2$ has both oscilating and constant terms .
Oscilating term corresponde to $t=s=t_{\pm}$ which give
order $N^{-1}$ for $\exp(-\frac{4\pi^2 m^2 N}{A}) w(m,-m)$,
and constant term correspondeto  $t=t_{\pm},s=t_{\mp}$ which give
order $N$ for $\exp(-\frac{4\pi^2 m^2 N}{A}) w(m,-m)$.
In the large $N$ limit we obtain
\begin{equation}
Z_{2 {\rm inst}}
              \simeq -\frac{2N}{\pi}[\quad\frac{\pi}{2} -
                             \arcsin(\frac{2\pi^2 m^2}{A}-1)-
                             \frac{2\pi^2 m^2}{A} \sqrt{\frac{A}{\pi^2 m^2}-1}
\quad] .
\end{equation}

\section{Numerical calculation }

In this section we show the result of numerical calculation of
 $\alpha_1,\alpha_2,\alpha_3,\alpha_4$ and free energies for $N$=3,4,5
using the {\sl Mathematica }.
We assume $0<A<4\pi^2$. From the above argument, in this region  we can
restrict the instanton charge as $m_{i}=\pm 1$
(which we call \lq truncated\rq )
 in multi-instanton sector.

Fig1,2 show that $N=10$, ${\sl a}\equiv A/\pi^2 =2$ results respectively for
 $\alpha_1,\alpha_2,
\alpha_3$ and $\alpha_4$.
We observe that compared to the $\alpha$ 's with ${\sl a}>1$ , the $\alpha$ 's
 with ${\sl a}<1$ are negligible.
Moreover we see remarkable difference between
$\alpha_1,\alpha_3 $ (even j) and $\alpha_2,\alpha_4$ (odd j)
with respect to  order and shape in the region of ${\sl a}>1$.
The shapes of $\alpha_2,\alpha_4$ are almost same and both have the
properties of non-oscilating and linear scaling with respect to $N$.
On the other hand the shapes of $\alpha_1,\alpha_3 $ are different.
But both have the properties of oscilating and $N^{-1/2}$ scaling .
We already observed that  $\alpha_1$
has $N^{-1/2}$ scaling and $ \alpha_2 $ has linear($N$) scaling using
 the large $N$ saddle point method.
The numerical calculation indicates the agreement with  eq (29),(42).
{}From this observation and the argument of section 2, we conjecture the
following scenario of the large $N$ phase transition.
That is, $\alpha_j$ with even $j$ have linear scaling with respect to $N$ in
the region of ${\sl a}>1$
and inside the exponential in eq (21), we have
\begin{equation}
\sum_{j=1}^{N} \frac{(-)^{j-1}}{j} z^{j} \alpha_{j} \simeq
-\sum_{n=1}^{[N/2]} \frac{1}{2n} z^{2n} \alpha_{2n} \simeq {\cal O}(N^2).
\end{equation}
This gives order $N^2$ result for the free energy in ${\sl a}>1$.
Thus $\alpha_j$ with even $j$ equally contrubute to the
large $N$ phase transition.

Fig3 show the truncated free energies
\begin{equation}
F=\frac{1}{N^2} \log [ \oint \frac{ dz }{2\pi i}
     \frac{1}{z^{N+1}}
     \frac{1}{1-z} \exp(\sum_{j=1}^{N} \frac{(-)^{j-1}}{j} z^{j}
     \alpha_{j} ) ]
\end{equation}
for $N$=3,4 and 5.
The partiton function for $N$=5 turns to have negative value in
${\sl a}{> \atop \simeq} 3.2$.
This causes the singular behavior around ${\sl a}\simeq 3.2$ of the $N=5$ graph
(here we plot real part of the $F$). We observe that the graphs stand
up at ${\sl a}\simeq 1$ and in the neighborhood of  ${\sl a}=1$ the
graphs seem to
converge in the large $N$ limit.
But in the region ${\sl a}{> \atop \simeq} 2.5$ discrepancy of the $N$=3,4
and 5 graphs is large. We think the truncation
becomes invalid in this region.
Hence we may conclude  that near $A=\pi^2$ nature of the large $N$
phase transition remain at $N=3,4,5$ level, but we
cannot distiguish whether
this behavior change is smoose or not.

\section{Outlook}
We have reformulated the multi-instanton amplitude as
 the multiple contour integrals and found that this represenation make clear
the properties and the effect of multi-instanton in two dimensional QCD.
In particular ``neutral'' configurations of the even number instantons
 are dominated in
the large $N$ limit.
But we think further investigation is needed about
\begin{enumerate}
\item derivation of the order of large $N$ phase transition
 in terms of our contour integral representation,
\item  clearcut order parameter and Ginzburg-Landau
 effective action,
\item  possibility of phase transitions at $A=\pi^2 k^2$ ($k$ :positive
integer).
\end{enumerate}

Some explanations are needed.

About point 1, we did not obtain the  free energy analytically because we do
not
know general method for solving the singular saddle point problem.
But if such method is found out, we will obtain analytic form of
the free energy
and hence the order of the  phase transition.

The reader may be confused  with resspect to point 2.
By \lq clearcut\rq , we intend  order parameter which is zero or nonzero
according to $A {> \atop <} \pi^2 $. In order to drive  argument of
phase transition to Ginzburg -Landau type, we must know such order
parameter a priori as  variable of effective action.
In our case instanton number is naive contender, but it is ruled out. From the
expression of instanton amplitudes they are  not positive definit and
can not have the meaning of probability\footnote{That is why we don't use the
word \lq instanton condensation\rq .} ( Boltzman weight).
Hence we must consider other order parameter.
If such order parameter is found out, the effective action will be the
following form near critical point \footnote{In the case of n th order
phase transition,$g \phi^3 $ term
is replaced by $g \phi^{2n/(n-1)} $. }.
\begin{equation}
\Gamma (\phi )= -(\beta-\beta_c )\phi^2 +g \phi^3 \qquad (g >0)
\end{equation}

About point 3, we ignore the instanton configuration of $m=\pm 2,\pm
3,\ldots $ in section 4 since such configuration give exponentioal
damping factor in the region of $\pi^2 <A<4\pi^2$. Therefore, same argument
leads to that in the region of $k^2 \pi^2 <A< (k+1)^2 \pi^2 $ ($k$ :
positive integer)
 we can ignore
the instanton configurations of $m=\pm (k+1),\pm (k+2),\ldots $. Hence the
multi-instanton configurations which should be included are changed
 at $A=k^2 \pi^2 $.
Other approaches did not predict phase transitions at $A=k^2 \pi^2 $,
but we think
whether the changes lead to some phase transitions or not is still important
 problem. \\

We hope this work will shed some lights on the global structure of two
 dimensional
 QCD.

\vspace{2cm}
\begin{center}
{\large {\bf Acknowledgements}}
\end{center}
\vspace{0.5cm}
The auther would like to thank Professor K.Ishikawa for his suggestion and
careful reading the manuscript.

\newpage
\leftline{{\Large {\bf Appendix \hspace {0.5cm}
  Multi-instanton effect in Wilson loop}}}
\vspace{0.5cm}
In this appendix  we note the multi-instanton effect in Wilson loop.
Gross and Matytsin also showed the formula for  Wilson loop on a sphere
 \cite{gm2}:
\begin{equation}
W_{n}(A_1,A_2)\equiv \langle \frac{1}{N} {\rm tr}U^{n} \rangle .
\end{equation}
Here the loop devide the sphere with area $A$ into two  disks with
area  $A_1,A_2$ and $U$ is gauge field holonomy along the loop .
\begin{eqnarray}
& & W_{n}(A_1,A_2) Z_{QCD} \nonumber \\
&=&\ee^{\frac{A}{24}(N^2-1)}(\frac{N}{A})^{N^2}
\frac{1}{N}\sum_{k=1}^{N} \sum_{\{m\}}
\ee^{-\frac{2\pi^2 N}{A}\sum_{i=1}^{N} m_{i}^{2}-2\pi i n m_{k}
         \frac{A_2}{A} -\frac{n^2 A_1 A_2}{2NA}} \nonumber \\
& \times & \int_{-\infty}^{\infty} \prod_{i=1}^{N} dy_{i}
                  \Delta (y_{i}+2\pi m_{i}+\frac{inA_2}{N}\delta_{ik})
                  \Delta (y_{i}-2\pi m_{i}+\frac{inA_1}{N}\delta_{ik})
                  \ee^{-\frac{N}{2A} \sum_{i=1}^{N} y_{i}^{2}} .
\end{eqnarray}
{}From above equation, calculating Wilson loop is almost same as
calculating the  partition function with additional imaginary charged
instanton.
Hence in the same way as the partition function,  we can rewrite this
as multiple contour integrals:
\begin{eqnarray}
&=& Z^{{\rm weak}} \sum_{l=0}^{N} \frac{1}{l!}
     \sum_{m_1,\ldots,m_l \neq 0}
     \ee^{-\frac{2\pi^2 N}{A}(m_1^2 +\cdots +m_l^2)}
     \oint \frac{ dt_{1} }{2\pi i} \cdots \oint \frac{ dt_{l} }{2\pi i}
     \nonumber \\
 & & \times  \ee^{-\frac{ 4 \pi^2 N }{A}(m_{1}^2  t_{1}+\cdots+
                    m_{l}^2 t_{l} ) }
                  (1+\frac{1}{t_{1}})^{N} \cdots (1+\frac{1}{t_{l}})^{N}
      \nonumber \\
  & & \times [\frac{l}{N} \ee^{-2\pi i n m_1 \frac{A_2}{A}
                                +2\pi i n m_1 \frac{A_1-A_2}{A} t_1
                                -\frac{n^2}{N}\frac{A_1 A_2}{A} t_1}
                        {\rm det}^{(l)}B \nonumber \\
      &  & \qquad +\frac{1}{N} \oint \frac{ dt_{l+1} }{2\pi i}
                        \ee^{-\frac{n^2}{N}\frac{A_1 A_2}{A}t_{l+1}}
                        (1+\frac{1}{t_{l+1}})^{N}
                        {\rm det}^{(l+1)}C  ] ,
\end{eqnarray}
where
\begin{eqnarray}
B_{ij}&=& \frac{1}{\frac{\beta_j}{\beta_i}(1+t_{j})-t_{i}} , \\
C_{ij}&=& \frac{1}{\frac{\gamma_j}{\gamma_i}(1+t_{j})-t_{i}}
\end{eqnarray}
and
\begin{eqnarray}
\beta_i &=&-2\pi m_i +\frac{in}{N}A_1 \delta_{i1} ,   \\
\gamma_i &=& -2\pi m_i \quad ({\rm for} \quad i=1,\ldots ,n) , \\
\gamma_{n+1} &=& \frac{in}{N}A_1 .
\end{eqnarray}
Using this formula we can calculate multi- instanton effect of
the Wilson loop in the large N limit.

For example in  0 instanton sector (weak coupling phase), we obtain,
\begin{equation}
W_{n}(A_1,A_2) Z_{QCD} |_{0 {\rm inst}}=
   Z^{{\rm weak}}\frac{1}{N}
   \oint \frac{ dt }{2\pi i}\ee^{-\frac{n^2 A_1 A_2}{NA}t}
   (1+\frac{1}{t})^{N} .
\end{equation}
By rescaling  $t \to Nt $, we obtain the known result:
\begin{equation}
  W_{n}(A_1,A_2) \simeq \frac{1}{n}\sqrt{\frac{A}{A_1 A_2}}J_{1}(2n
     \sqrt{\frac{A_1 A_2}{A}}) .
\end{equation}

\newpage
\centerline{{\Large {\bf Figure captions}}}
\vspace{0.5cm}
\noindent
Fig 1: The truncated $\alpha$ 's for $ N=10 $ and various  ${\sl a}
\equiv A/\pi^2 $ are shown.  \\
Fig 2: The truncated $\alpha $ 's for ${\sl a}=2 $ and various $ N $
         are shown. \\
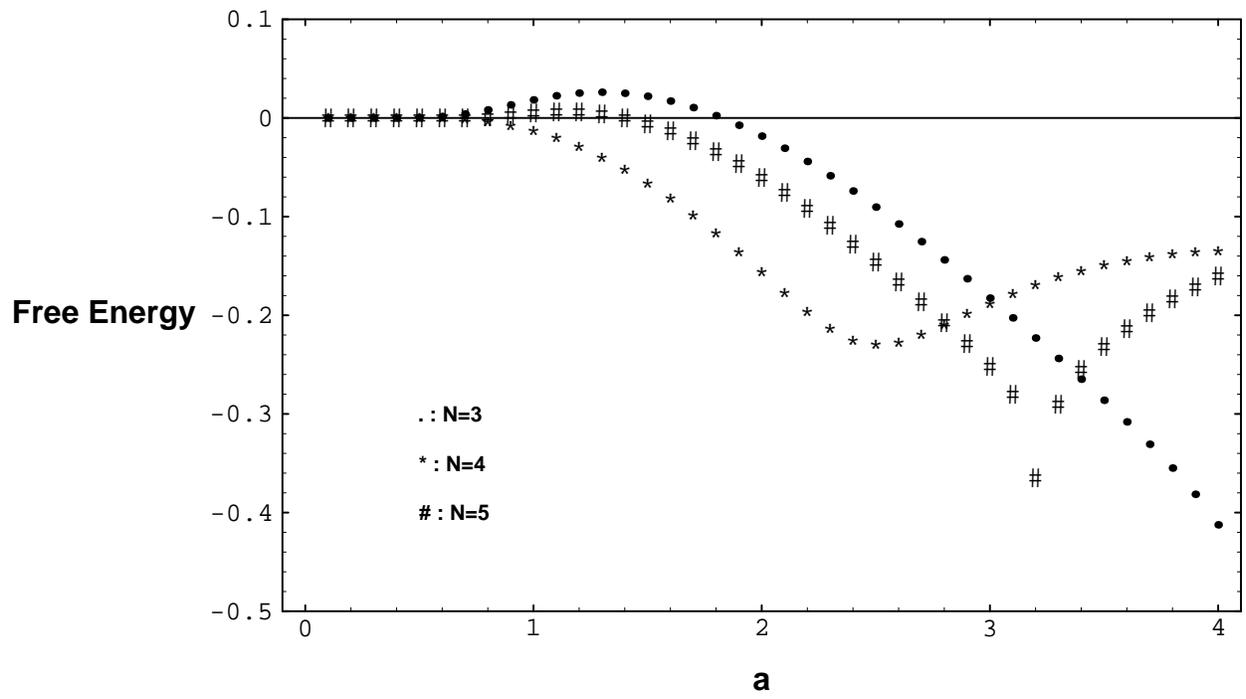
Fig 3: The truncated free energies for $N=3,4,5$ and various  ${\sl a}$
are shown.


\begin{thebibliography}{99}
 \bibitem{gt}
     D.Gross , Nucl.Phys.{\bf B400}(1993) 161; \\
     D.Gross and W.Taylor , Nucl.Phys.{\bf B400} (1993) 181; \\
     D.Gross and W.Taylor , Nucl.Phys.{\bf B403} (1993) 395.
 \bibitem{dk}
    M.Douglas and V.Kazakov , Phys.Lett.{\bf B319} (1993) 219.
 \bibitem{w}
     E.Witten, Commun. Math. Phys. {\bf 141} (1991) 153 ;
                J. Geom. Phys. {\bf 9} (1992) 303.
 \bibitem{mp}
    A.Minahan and Polychronakos , Nucl.Phys.{\bf B422} (1994) 172.
 \bibitem{gm1}
     D.Gross and A.Matytsin , Nucl.Phys.{\bf B429} (1994) 50.
 \bibitem{cdmp}
     M.Caselle, A.D'Adda, L.Magnea and S.Panzeri , in: Proc. 1993
             Trieste Workshop on High Energy Physics and Cosmology.
 \bibitem{gm2}
     D.Gross and A.Matytsin , hep-th/9410054.
 \bibitem{t}
     W.Taylor , hep-th/9404175.
 \bibitem{mr}
    A.Migdal , Sov.Phys.JETP {\bf 42} (1975) 413; \\
    B.Rusakov , Mod.Phys.Lett. {\bf A5} (1990) 693.
\end{thebibliography}
\end{document}